\documentclass[conference]{IEEEtran}
\IEEEoverridecommandlockouts
\usepackage{cite}
\usepackage{amsmath,amssymb,amsfonts}
\usepackage{graphicx}
\usepackage{textcomp}
\usepackage{xcolor}
\def\BibTeX{{\rm B\kern-.05em{\sc i\kern-.025em b}\kern-.08em
		T\kern-.1667em\lower.7ex\hbox{E}\kern-.125emX}}

\usepackage{graphicx,cite,calc,color,subfigmat,amssymb,amsmath,dsfont,epstopdf,tikz}
\usepackage{array}
\usepackage{multirow}
\usepackage{cleveref}

\usepackage{subfigure}
\usepackage{graphicx}
\usepackage{amsmath,amssymb,amsfonts,bm}
\usepackage{breqn}

\usepackage{algorithm}
\usepackage[noend]{algpseudocode}

\makeatletter
\def\BState{\State\hskip-\ALG@thistlm}
\makeatother

\usepackage{graphicx}
\usepackage{caption}

\usepackage{tabularx}
\usepackage{dblfloatfix}

\usepackage{fancyhdr}
\usepackage{kantlipsum}

\usepackage{etoolbox}

\begin{document}
	
	\title{Content-based image retrieval speedup\\
	}
	
	\author{\IEEEauthorblockN{Sadegh Fadaei}
		\IEEEauthorblockA{\textit{Department of Electrical Engineering } \\
			\textit{Faculty of Engineering}\\
			\textit{Yasouj University}\\
			Yasouj, Iran, 7591874934 \\
			s.fadaei@yu.ac.ir}
		\and
		\IEEEauthorblockN{Abdolreza Rashno}
		\IEEEauthorblockA{\textit{Department of Computer Engineering } \\
			\textit{Engineering Faculty}\\
			\textit{Lorestan University}\\
			Khorramabad, Iran \\
			rashno.a@lu.ac.ir}
		\and
		\IEEEauthorblockN{Elyas Rashno}
		\IEEEauthorblockA{\textit{Department of Computer Engineering} \\
			\textit{Iran University Of Science and Technology}\\
			Tehran, Iran \\
			elyas.rashno@gmail.com}

	}
	
	\maketitle

\begin{abstract}

Content-based image retrieval (CBIR) is a task of retrieving images from their contents. Since retrieval process is a time-consuming task in large image databases, acceleration methods can be very useful. This paper presents a novel method to speed up CBIR systems. In the proposed method, first Zernike moments are extracted from query image and an interval is calculated for that query. Images in database which are out of the interval are ignored in retrieval process. Therefore, a database reduction occurs before retrieval which leads to speed up. It is shown that in reduced database, relevant images to query image are preserved and irrelevant images are throwed away. Therefore, the proposed method speed up retrieval process and preserve CBIR accuracy, simultaneously.    
\end{abstract}


\section{Introduction}
Image processing algorithms have been widely used in many applications such as pixel classification \cite{rashno2017effectiveNeuCom,rashno2015MarsConf}, image restoration \cite{rashno2014imageImgRes,rashno2014regularizationImgRes}, medical image analysis \cite{r101,r111,r121,r131,r141,r151,r161,r171}. Image processing applications are using data mining and machine learning algorithms like other applications such as speech processing methodologies \cite{rashno2013highlySpeVer,rashno2015textSpeVerConf} and neutrosophic data analysis\cite{r24,Elyas2018EAAI,Elyas2019SpeechConv,ElyasICEE2019certainty,ElyasKBEI2019}.
Content-based image retrieval (CBIR) is retrieving images based on their structural and conceptual characteristics without needing for manual annotation. CBIR can be done with image processing methodologies or machine learning and data mining algorithms. CBIR is among important fields in image processing with applications including art collections, crime prevention, architectural and engineering design, geographical information, intellectual property, medical diagnosis, face finding, textiles industry, photograph archives, military, retail catalogs and remote sensing systems\cite{r1}. Feature extraction and retrieval are two main steps in CBIR. For large databases, retrieval phase is very time-consuming especially for cases with complicated measures for image similarity evaluation. Due to the increasing the size of image databases, research in the field of CBIR systems has gained high attention\cite{r2}. 

Many CBIR systems have been proposed in recent years. In \cite{r3}, a fast CBIR system based on a fusion of weighted color, texture, chromaticity moments, color percentile, and local binary pattern (LBP) was proposed followed by a supervised query image classification and retrieval model. 
In \cite{r4}, a normalization method as linear scaling to unit variance was proposed for  feature vector equalization. Based on this method, equal-average K nearest neighbor search (EKNNS) method was used to find the first K nearest neighbors of the query. These neighbors were considered as retrieved images.
In \cite{r5}, a CBIR system containing two steps image indexation and image retrieving was proposed. In the first step, MapReduce distributed model on Spark and Tachyon memory-centric distributed storage system were used to speed up the indexation task. In the second step, parallel k-Nearest Neighbors (k-NN) MapReduce model was implemented under Apache Spark. 
In \cite{r6}, a plane semantic ball (PSB) in graphics processing unit (GPU) was developed as adaptive index structure which could reduce the computational tasks of retrieval process parallel accelerated scheme. In PSB, a collection of pivots are used for representing semantics as well as multiple balls to cover more informative reference features in image which all are executed on GPU efciently.
In \cite{r7}, a method for CBIR acceleration was proposed on the RDISK machine. This machine uses the philosophy of smart-disks and is considered as a cluster of FPGA-enhanced hard-drives. The proposed platform combined coarse and fine grain parallelism in cluster nodes and programmable logic devices.

This work extends our previous researches in\cite{r10,fadaei2017local,rashno2017refined,rashno2019content,19_rashno2017content,rashno2015efficient} which was CBIR systems with different features and retrieval schemes. Here, the focus is speeding up the retrieval process. For this task, Zernike moments as shape features are selected and then an interval for each query image is computed which leads to exclude many irrelevant images and reduce the database in retrieval phase significantly. Although this method is proposed for Zernike features, it can be easily adapted for retrieval processes based on any feature type including color, texture and shape. The rest of this paper is organized as follows: Section II presents the proposed method. Experimental results and conclusion are discussed in sections III and IV, respectively.

\section{Proposed method}
\begin{figure}[h]
	\centering
	\captionsetup{justification=centering}
	\includegraphics[width= 0.5 \textwidth ]{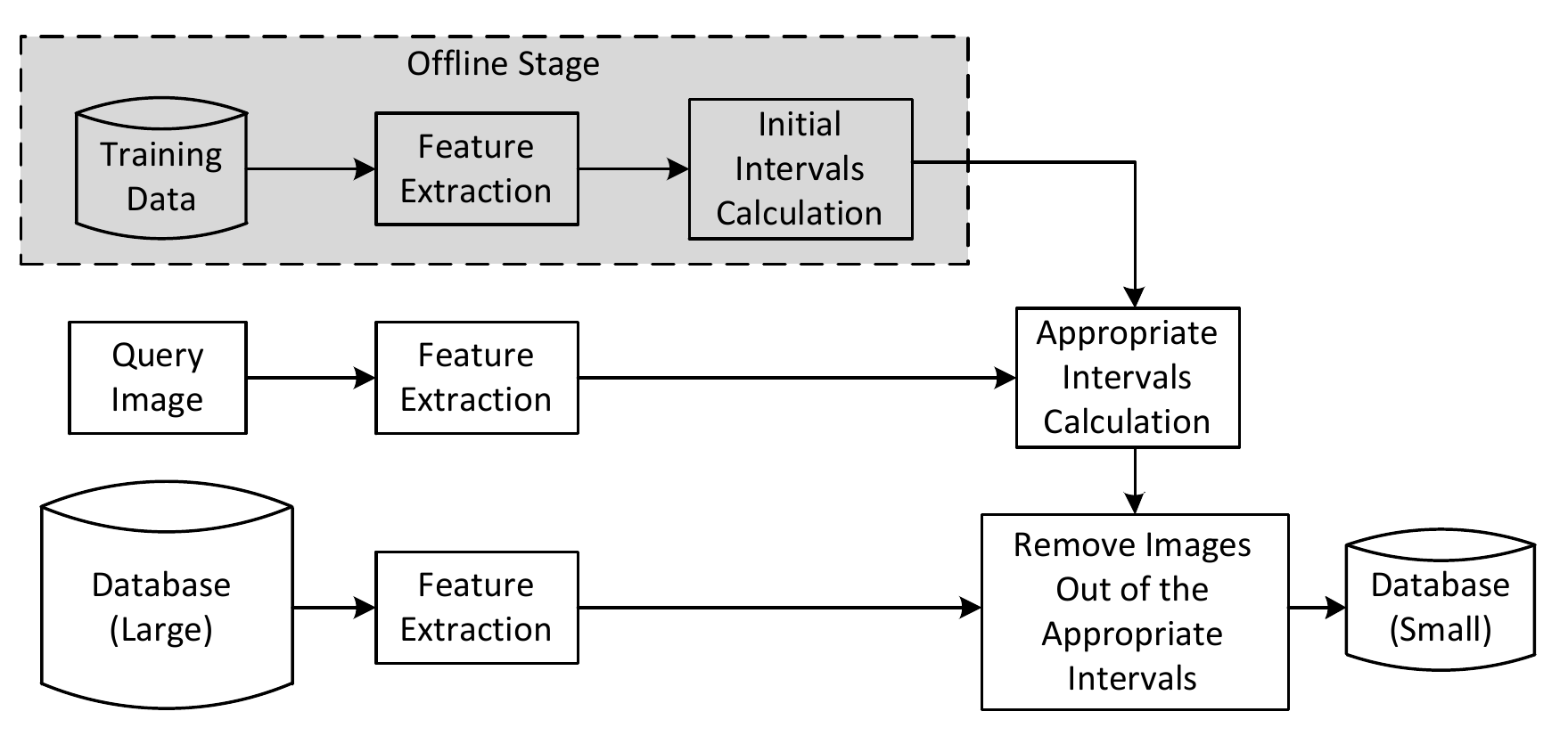}
	\caption{Flowchart of the proposed method.}
	\label{Flowchart}
\end{figure}

Feature extraction is a time consuming task in CBIR systems. Since features should be extracted from all images in database, this challenge rise for large databases. The main contribution of this work is to propose a method for quick eliminating of images which are irrelevant to query image. In fact, for each image query, this step is done and reduces database size significantly.  For this task, a few number of features are extracted from query image and optimum intervals are calculated. All images out of these intervals are ignored in retrieval phase. Note that extraction of considered feature for database reduction should be as quick as possible. The flowchart of the proposed method is shown in Fig. \ref{Flowchart}.

In this work, Zernike moments are used. These moments are derived from orthogonal Zernike polynomials and are a set of orthogonal moments[1]. Pseudo Zernike moments are an extension of Zernike moments with useful and important features including capabilities of multi-level representation.These moments are robust against noise and image rotation. Implementation of such features are easy and fast\cite{r8}  which is very important for our proposed method. Orthogonal polynomials $V_{mn}(x,y)$ can be represented as follows:

\begin{equation}
V_{mn}(x,y)=V_{mn}(\rho,\theta)=R_{mn}(\rho)\cdot e^{im\theta}
\label{eq1}
\end{equation}

\begin{equation}
R_{mn}(\rho)=\sum_{s=0}^{m-|m|}\frac{\displaystyle{(-1)^s(2m+1-s)!\rho^{m-s}}}{\displaystyle{s!(m+|n|+1-s)!(m-|n|-s)!}}
\label{eq2}
\end{equation}

By considering Orthogonality property of polynomials $f(x,y)$ can be decomposed as follows: 

\begin{equation}
f(x,y)=\sum_{m=0}^{\infty}\sum_{\{n:|n|\leq m\}} A_{mn}V_{mn}(x,y)
\label{eq3}
\end{equation}

\begin{equation}
A_{mn}=\frac{m+1}{\pi} \iint_{x^2+y^2\leq 1} f(x,y)V_{mn}^{\ast}(x,y)dxdy
\label{eq4}
\end{equation}

where $A_{mn}$ coefficients describe Pseudo ZMs and their size are used as image features[9]. 
In the proposed scheme for databse reduction, it is supposed that  there are $N$ image categories with $M$ image per category. For image $j$ in category $i$, one feature is extracted and represented by $f_{ij}$. Then, two parameters $a_i$ and $b_i$ are computed from $f_{ij}$ by Eq. \ref{eq5}.  

\begin{equation} \begin{array}{rl}
a_i=\text{min}(f_{i1},f_{i2},\dots,f_{iM}) \\
b_i=\text{max}(f_{i1},f_{i2},\dots,f_{iM}) \\
\end{array} \Longrightarrow I_i=[a_i,b_i]
\label{eq5}
\end{equation}

\begin{figure*}[h]
	\centering
	\captionsetup{justification=centering}
	\includegraphics[width=  \textwidth ]{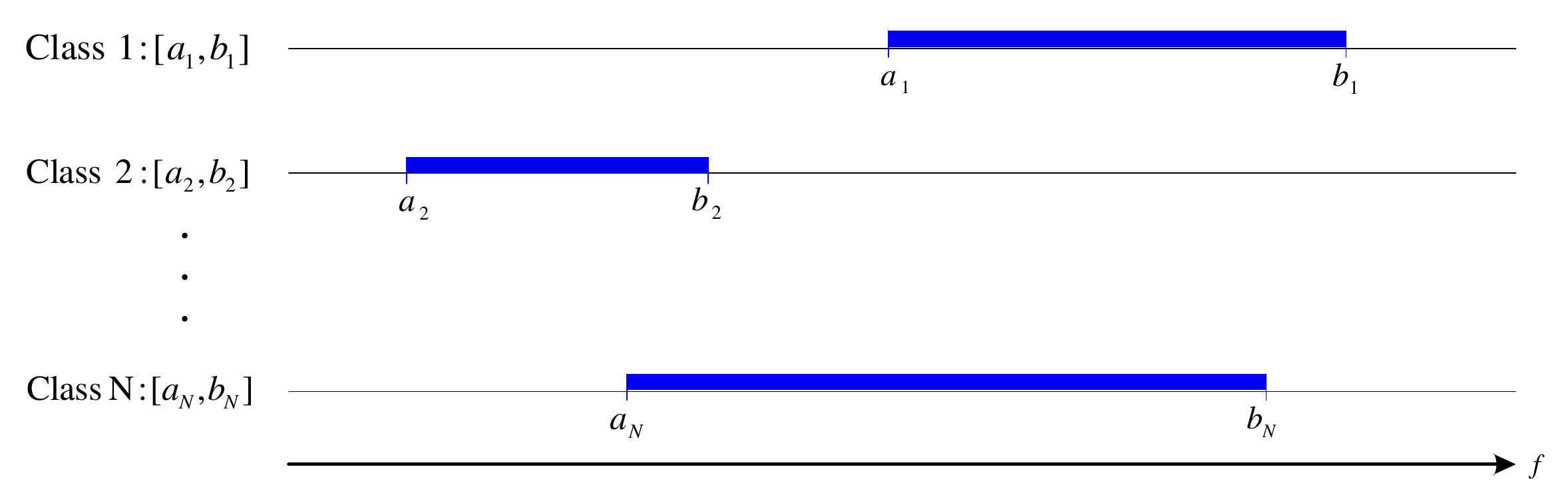}
	\caption{Bounds of feature in each category.}
	\label{FeatureBound}
\end{figure*}

Interval $I_i=[a_i,b_i]$ represents bounds of feature for category $i$ as shown in Fig. \ref{FeatureBound}

In the next step, a center and radius is calculated for each class as follows:

\begin{equation}
\text{Class } i \text{: } R_i=\frac{b_i-a_i}{2} \quad , \quad C_i=\frac{a_i+b_i}{2}
\label{eq6} 
\end{equation}

Then, $R_{max}$ is calculated from radius of classes:

\begin{equation}
R_{max}=\text{max}\{R_1,R_2,\dots,R_N\}
\label{eq7}
\end{equation}

Therefore, $C_1,C_2,...,C_N, R_{max}$ are computed in offline phase. In online step, by presenting query image with feature $f_q$, the following interval are calculated from data in offline step.

\begin{equation}
S_1=[f_q-R_{max},f_q+R_{max}]
\label{eq8}
\end{equation}

\begin{equation}
\begin{array}{rl}
S_2=[a_m,b_m] \quad , \quad m=\text{argmin}|f_q-C_i| \\
i \quad \quad \quad \quad \quad \\
\end{array}
\label{eq9}
\end{equation}

The final interval to select relevant images from database is computed by Eq. \ref{eq10}.

\begin{equation}
S_q=S_1 \cup S_2
\label{eq10}
\end{equation}

Therefore, after $S_q$ computation, all images out of this interval are ignored and images in this interval are forwarded for retrieval step. 

\section{Experimental Results}
To show the effectiveness of the proposed method, it is evaluated on Corel-1k database which contains 1000 images with 10 categories and 100 image in each category. Here, Zernike features $A_{00}$, $A_{20}$ and $A_{22}$ are used. First, it is shown that how much the database is reduced by the proposed method. Then, CBIR system is implemented on both the full-length database and reduced database and retrieval accuracies of two cases are compared. In all experiments, 50\% of database is used for train and remaind 50\% for test.

\subsection{Database reduction}

In the first step, the effect of each Zernike features as well as their combination are assessed for database reduction. Table I reports values for $C_1,C_2,...,C_N, R_{max}$ obtained in offline phase. 

\begin{table*}[h]
	\footnotesize
	\centering
	\caption{$C_1,C_2,\dots,C_N,R_{max}$ for Zernike features and Corel-1k database.}
	\label{Table1}
	\begin{tabular}{|l|l|l|l|l|l|l|l|l|l|l|l|}
		\hline
		& $C_1$  & $C_2$ & $C_3$ & $C_4$  & $C_5$   & $C_6$  & $C_7$ & $C_8$  & $C_9$ & $C_{10}$ & $R_{max}$  \\ \hline
		$A_{00}$ & $50.41$ & $41.65$ & $40.72$ & $38.76$ & $50.90$ & $81.03$ & $43.75$ & $32.95$ & $43.75$ & $41.45$ & $24.91$ \\ \hline
		$A_{20}$ & $24.09$ &	$41.45$ &	$44.87$ &	$31.03$ &	$33.51$ &	$78.92$ &	$51.60$ &	$18.38$ &	$44.73$ &	$44.63$ &	$40.62$ \\
		\hline
		$A_{22}$ & $26.63$ &	$32.68$ &	$50.90$ &	$33.86$ &	$36.83$ &	$75.64$ &	$41.38$ &	$15.26$ &	$29.67$ &	$46.45$ &	$36.56$ \\
		\hline
	\end{tabular}
\end{table*}

\begin{figure*}[!h]
	\centering
	\captionsetup{justification=centering}
	\includegraphics[width=  \textwidth ]{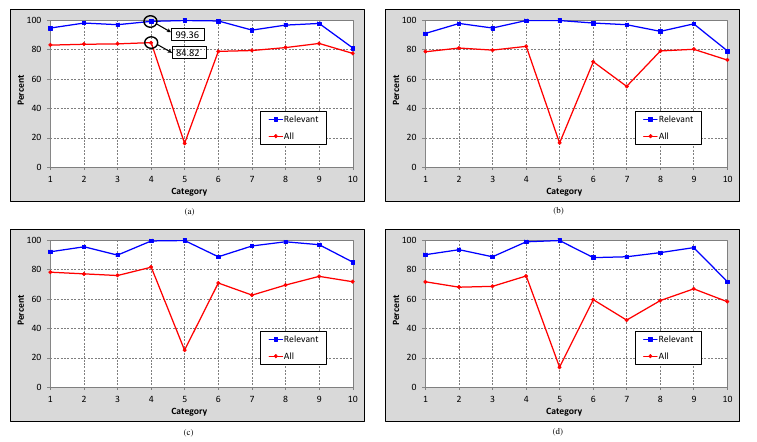}
	\caption{Bars for remained percentage of relevant images and all images retrived by Zernik feature: (a)$A_{00}$, (b)$A_{20}$, (c)$A_{22}$ and (d) all features .}
	\label{RedecedDataset}
\end{figure*}

In online phase, a query image is presented and database is reduced to a new database. Then, number of relevant images remained in new database is calculated. Also, reduction percent of new database in comparison with the main database is calculated. This experiment is done for all query images in test set (500 images) and then average remained percentage is depicted in Fig. \ref{RedecedDataset}. In Corel-1k database, it is expected that remained percentage be as close as 100\% for relevant images with query image  and as close as 10\% for the whole images. As it is clear from Fig. \ref{RedecedDataset}, almost all relevant images are kept and blue line is near to 100\% and meets 10\% for the all images in some categories.  For instance, in Fig. \ref{RedecedDataset}(a), remained percentage in category 4 is 99.36\% and 84.82\% for all images and relevant images, respectively.

\begin{figure}[!h]
	\centering
	\captionsetup{justification=centering}
	\includegraphics[width= 0.5 \textwidth ]{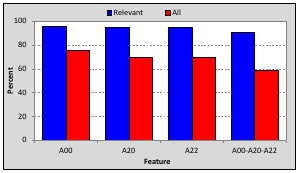}
	\caption{Percentage of remained relevant images and all images with Zernike features.}
	\label{RetAccAll}
\end{figure}

The percentage of remained relevant images and remained all images for each Zernike feature and all features are depicted in Fig. \ref{RetAccAll}. It is clear that by all Zernike features, remained percentage of relevant images is 91.17\% and remained percentage of all images is 59.16\%. It means the the majority part of irrelevant images are throwed away while a high percentage of relevant images is preserved.

\subsection{Retrieval time after database reduction}
\begin{figure}[h]
	\centering
	\captionsetup{justification=centering}
	\includegraphics[width= 0.5 \textwidth ]{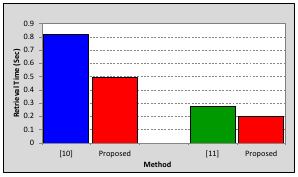}
	\caption{Retrival time with and without database reduction.}
	\label{RetTime}
\end{figure}

Database reduction leads to speed up in retrieval phase since retrieving images from large databases needs more time than small databases. To show that how reduced database obtained from the proposed method speed up the retrieval process, two CBIR methods proposed in\cite{r10,r11} are implemented and evaluated by the whole database and reduced database. Retrieval time for these cases are shown in Fig. \ref{RetTime}.

\subsection{Retrieval accuracy after database reduction}
\begin{figure}[h]
	\centering
	\captionsetup{justification=centering}
	\includegraphics[width= 0.5 \textwidth ]{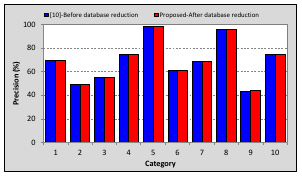}
	\caption{Retrival accuracy of method [10] before and after database reduction.}
	\label{RetAcc}
\end{figure}

\begin{figure}[h]
	\centering
	\captionsetup{justification=centering}
	\includegraphics[width= 0.5 \textwidth ]{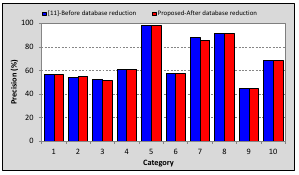}
	\caption{Retrival accuracy of method [11] before and after database reduction.}
	\label{RetAcc1}
\end{figure}

By reducing the database size, the accuracy of CBIR system is decreased since relevant images may be removed in database reduction phase. The last experiment is done to evaluate CBIR accuracy before and after database reduction by the proposed method. For this task, the accuracy of methods in \cite{r10,r11} are compared before and after database reduction in Figs. \ref{RetAcc} and \ref{RetAcc1}, respectively. It shows that database reduction does not decrease CBIR system accuracy since the majority part of relevant images are also in reduced database. In this experiment, average precisions for method \cite{r10} are 69.26\% and 69.20\%, respectively. These precisions for method\cite{r11} are 67.50\% and 67.26\% which means that precisions are preserved after database reduction.   

\section{Conclusion}

This paper presented a fast and efficient method to speed up CBIR system. The proposed method was based on an interval extracted from Zernike moments which excluded all images out of this interval before retrieval process. Experiments on Corel-1k database showed that the proposed scheme decreases retrieval time significantly with the same retrieval accuracy in comparison with existing CBIR systems. Future efforts will be directed towards proposing speed up methods with efficient intervals extracted from texture and color features which are good descriptors for images with irregular objects. Finally, an optimize combination of intervals from different features can be proposed as another future work.

\end{document}